\documentclass{article}

\usepackage{PRIMEarxiv}
\usepackage{amssymb}
\usepackage{hyperref}
\usepackage[utf8]{inputenc} 
\usepackage[T1]{fontenc}    
\usepackage{hyperref}       
\usepackage{url}            
\usepackage{booktabs}       
\usepackage{amsfonts}       
\usepackage{nicefrac}       
\usepackage{microtype}      
\usepackage{lipsum}
\usepackage{fancyhdr}       
\usepackage{graphicx}       
\graphicspath{{media/}}     

\title{Mpox-AISM : AI-Mediated Super Monitoring for Mpox and Like-Mpox	
}

\author{
  Yubiao Yue \\
  School of Biomedical Engineering \\
  Guangzhou Medical University \\
  Guangzhou\\
  \texttt{jiche2020@126.com} \\
   \And
 Minghua Jiang \\
  Dermatological department \\
  Foshan Sanshui District People's Hospital \\
  Foshan\\
  \texttt{Jmh5270@sina.com} \\
  \AND
 Xinyue Zhang \\
  School of Biomedical Engineering \\
  Guangzhou Medical University \\
  Guangzhou\\
  \texttt{moonkkaabc@163.com} \\
  \And
 Jialong Xu \\
  School of Biomedical Engineering \\
  Guangzhou Medical University \\
  Guangzhou\\
  \texttt{jialong\_xu18@163.com} \\
  \And
 Huacong Ye \\
  School of Biomedical Engineering \\
  Guangzhou Medical University \\
  Guangzhou\\
  \texttt{davis\_yhcgpnu@yeah.net} \\
    \And
 Fan zhang \\
  Department of science and education \\
  Foshan Sanshui District People's Hospital \\
  Foshan\\
  \texttt{zhangfanyisheng\_01@163.com} \\
    \And
 Zhenzhang Li \\
  School of Mathematics and Systems Science \\
  Guangdong Polytechnic Normal University \\
  Guangzhou\\
  \texttt{zhenzhangli@gpnu.edu.cn} \\
    \And
 Yang Li \\
  School of Biomedical Engineering \\
  Guangzhou Medical University \\
  Guangzhou\\
  \texttt{lychris@sina.com} \\
}

\begin{document}
\maketitle

\begin{abstract}
Swift and accurate diagnosis for earlier-stage monkeypox (mpox) patients is crucial to avoiding its spread. However, the similarities between common skin disorders and mpox and the need for professional diagnosis unavoidably impaired the diagnosis of earlier-stage mpox patients and contributed to mpox outbreak. To address the challenge, we proposed “Super Monitoring”, a real-time visualization technique employing artificial intelligence (AI) and Internet technology to diagnose earlier-stage mpox cheaply, conveniently, and quickly. Concretely, AI-mediated “Super Monitoring” (mpox-AISM) integrates deep learning models, data augmentation, self-supervised learning, and cloud services. According to publicly accessible datasets, mpox-AISM’s Precision, Recall, Specificity, and F1-score in diagnosing mpox reach 99.3\%, 94.1\%, 99.9\%, and 96.6\%, respectively, and it achieves 94.51\% accuracy in diagnosing mpox, six like-mpox skin disorders, and normal skin. With the Internet and communication terminal, mpox-AISM has the potential to perform real-time and accurate diagnosis for earlier-stage mpox in real-world scenarios, thereby preventing mpox outbreak.
\end{abstract}


\section{Introduction}
Mpox, also known as monkeypox, is an infectious disease caused by the mpox virus \cite{ref1,ref2}. In 2022, the World Health Organization declared the global Mpox outbreak a public health emergency of international concern. Since May 2022, over 90000 cases of monkeypox have been reported worldwide \cite{ref3}. Given the increasing number of cases of monkeypox, it is urgent to quickly and timely diagnose the monkeypox virus. Polymerase Chain Reaction (PCR) is the primary clinical technology for mpox diagnosis \cite{ref4}. Although its diagnosis result is credible, it is still not the best solution to forestall mpox spread. This is because the clinical features of mpox patients initially show symptoms resembling the flu \cite{ref5}, followed by skin rashes, which first appear on the face and then gradually spread to the extremities \cite{ref6}. From the appearance, symptoms of these rashes are easily confused with measles, chickenpox, eczema and other skin diseases \cite{ref7,ref8}. Furthermore, mpox symptoms usually start within three weeks of exposure to the virus, yet the mpox virus spreads to others from the time symptoms begin or even during the incubation period \cite{ref9}. When patients actively seek testing, they are usually in the later stage of conditions, which may have caused the widespread spread of the virus among the public.

The above facts demonstrate that employing traditional diagnostic techniques to diagnose late-stage mpox is not conducive to preventing mpox spread, particularly at crucial places like border customs or in high-risk areas with dense crowds such as hospitals and schools. Notably, PCR technology necessitates specialized operating equipment, skilled medical personnel, and significant costs. For areas with limited medical resources, such as the wild and rural areas in underdeveloped regions, timely and efficient mpox testing is often inaccessible. Meanwhile, the diagnostic results of PCR are often incorrect since the viremia lasts for a short span of time in relation to the time specimens are generally collected after symptoms begin \cite{ref8}. Furthermore, earlier-stage mpox rash resembles common skin diseases, frequently resulting in misdiagnosis and wrong treatment. Even more worryingly, most of the cases of Mpox outbreak have been reported in locations that have historically been free from the infection \cite{ref7}, thereby highlighting the fact that developing an easy-to-use, low-cost, rapid and accurate method for the diagnosis of earlier-stage mpox is critically significant for swiftly preventing the spread of the outbreak and avoiding the deterioration of patients' physical conditions \cite{ref8}.

Given the above issues, artificial intelligence (AI) offers a reliable solution \cite{ref10}. Recently, deep learning (DL) models in AI have achieved significant success in various fields, including machine vision, medical imaging, and driverless vehicles. They have become sophisticated tools for addressing previously unknown and challenging problems. Moreover, AI-based innovations have reduced the cost of expensive equipment and professional knowledge required in the clinical diagnosis process, thereby helping overcome resource-constrained environments \cite{ref11}. In this work, we employed the Internet, popular DL algorithms, data augmentation and a novel self-supervised learning (SSL) strategy, i.e. "A Simple Framework for Contrastive Learning of Visual Representations" (SimCLR) \cite{ref12}, to design a real-time and online strategy for diagnosing mpox. Our strategy considers the clinical characteristics of mpox rash and its high similarity to common skin diseases and helps diagnose earlier-stage mpox conveniently and swiftly. We validated the effectiveness of our strategy using publicly available datasets. Furthermore, we classified the mpox rash images into four grades (I, II, III, and Others) and two stages (earlier and later) based on dataset characteristics and the clinical evolution trend of the mpox rash. In detail, Grade-I means body parts are more prone to pox rash, i.e., the faces, necks, and hands \cite{ref13,ref14}. At the same time, Grade-II contains the arms and legs, which are not easily covered by clothing and other articles of daily use. We retested our strategy and found that it still performs well in diagnosing mpox images in different parts and stages. Particularly, to improve the reliability and security of our strategy, we also visualized and explained its decision-making process and results.

\begin{figure}
    \centering
    \includegraphics[width=1\textwidth]{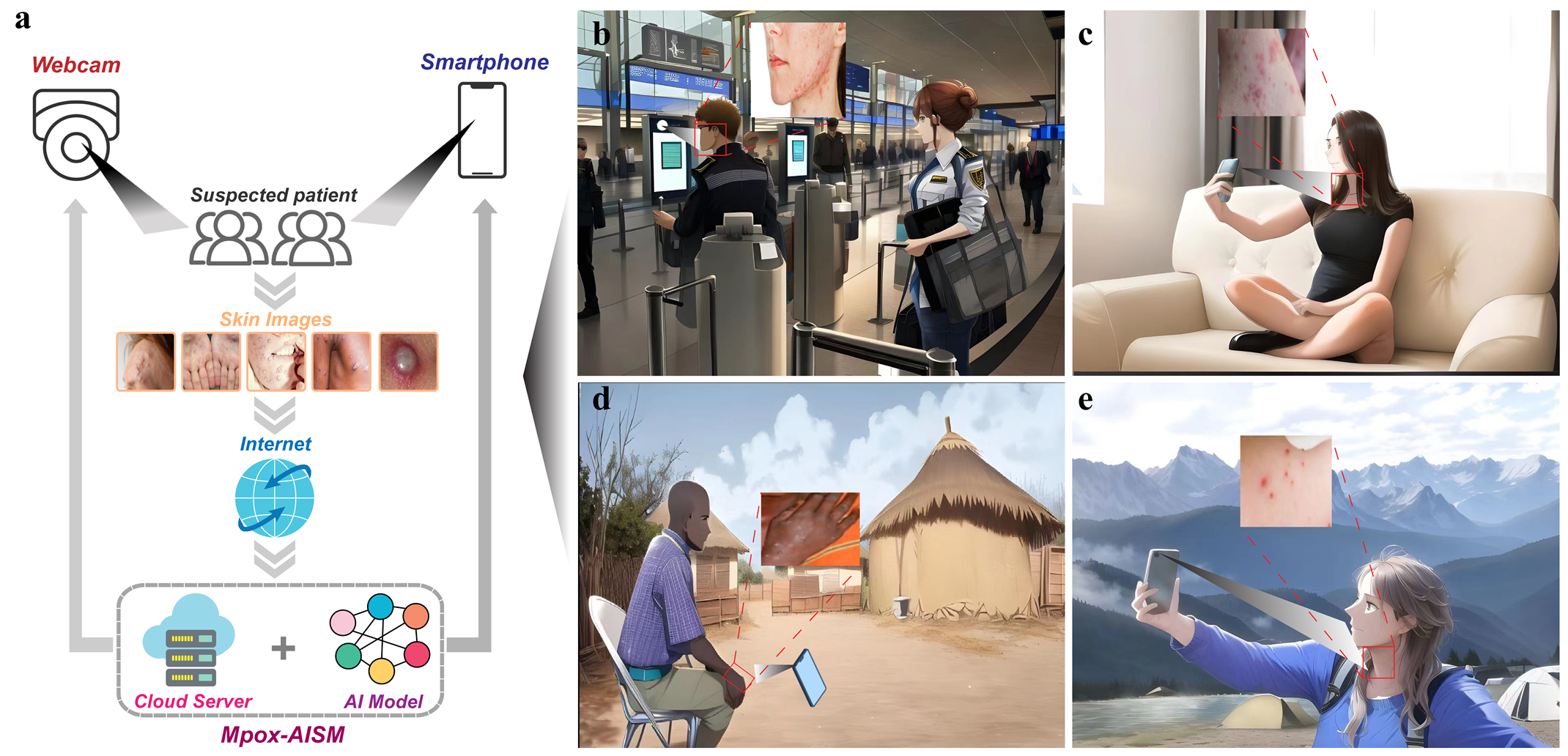}
    \caption{Application process of Mpox-AISM. b-e: Application settings in different situations.}
    \label{fig1}
\end{figure}

Based on the advantages above, we have named this strategy AI-mediated “Super Monitoring” (Abbreviated as Mpox-AISM, Fig. \ref{fig1}). Mpox-AISM has the potential to diagnose earlier-stage mpox in various settings, such as entry-exit inspection in airports and customs (Fig. 1b), family doctors (Fig. 1c), a rural area in an underdeveloped region (Fig. 1d), the wild and other settings (Fig. 1e). By deploying mpox-AISM in these settings, users only need to capture images of human skin using networked devices with lenses and upload them to our cloud server via the Internet, and then Mpox-AISM will then return the results to the user in real-time (Fig. 1a). In the end of the work, we have also developed a smartphone application based on Mpox-AISM (Mpox-AISM App) to provide free mpox testing services to the public and healthcare providers.

\section{Results}
\subsection{Mpox-AISM Design Workflows}
In this study, we selected ten classical classification models from the field of computer vision, namely VGG-19, GoogleNet, ResNet101, ResNeXt101\_64x4D, DenseNet201, EfficientNet\_B0, RegNetY\_16GF, RegNetX\_32GF, Vision Transformer Base, and Swin Transformer Base. Before training and testing these models, we expanded Data\_A into Data\_C through data augmentation. Subsequently, Data\_C was used to train and test each model under identical experimental conditions. Owing to the requirements for SSL, the EfficientNet-B0 and ResNeXt-101\_64x4D models were selected as final candidate models based on their superior performance. To further improve model performance, SimCLR and the SSL Dataset were employed to pre-train these two models. Following SSL, Data\_C was utilized to retrain and retest the pre-trained models. Experimental results reveal that ResNeXt101\_64x4D with SimCLR achieved state-of-the-art (SOTA) performance. The SOTA model was ultimately deployed to the cloud server, and Mpox-AISM was constructed. Mpox-AISM is capable of interacting with smartphones, webcams, and other network devices via the Internet. Users need only capture images of the lesioned skin area using the lenses of networked devices and upload them to Mpox-AISM. Subsequently, a real-time primary diagnosis result will be provided. The specific workflow is shown in Fig. 2.
\begin{figure}
    \centering
    \includegraphics[width=1\textwidth]{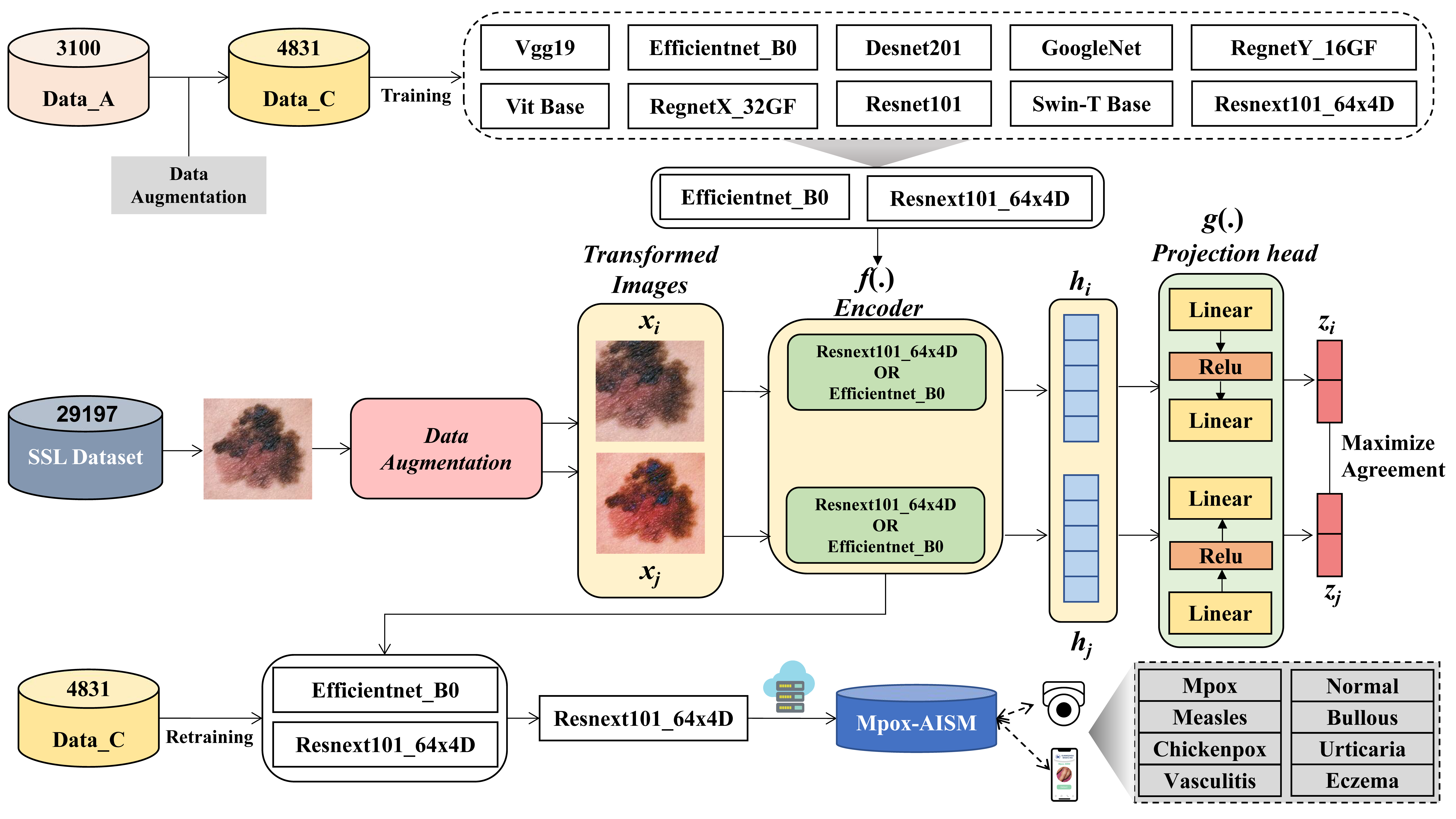}
    \caption{The workflow of this study.}
    \label{fig2}
\end{figure}
\subsection{Model Screening, Self-Supervised Learning, Retraining and Retesting}
The Test Accuracy and Train Loss were used to evaluate the candidate models. The Test Accuracy and Train loss of each candidate model were recorded during 300 epochs (Fig. 3a). Among all models, EfficentNet\_B0 had the highest Accuracy (90·57\%, 273th epoch), followed by Resnext101\_64x4D (84·97\%, 258th epoch), and their trends were also similar in terms of Train loss (Fig. 3b). Typically, higher Accuracy means better performance. However, due to the use of SSL, the structure of the model architecture was also considered. Like supervised learning, SSL benefits from deeper and wider networks7, which means that although the Accuracy of Resnext101\_64x4D is not as good as EfficientNet\_B0, Resnext101\_64x4D with SimCLR may outperform EfficientNet\_B0 with SimCLR due to its deeper and wider model architecture. Considering this fact, we chose these two models as the encoders of SimCLR.
\begin{figure}
    \centering
    \includegraphics[width=0.9\textwidth]{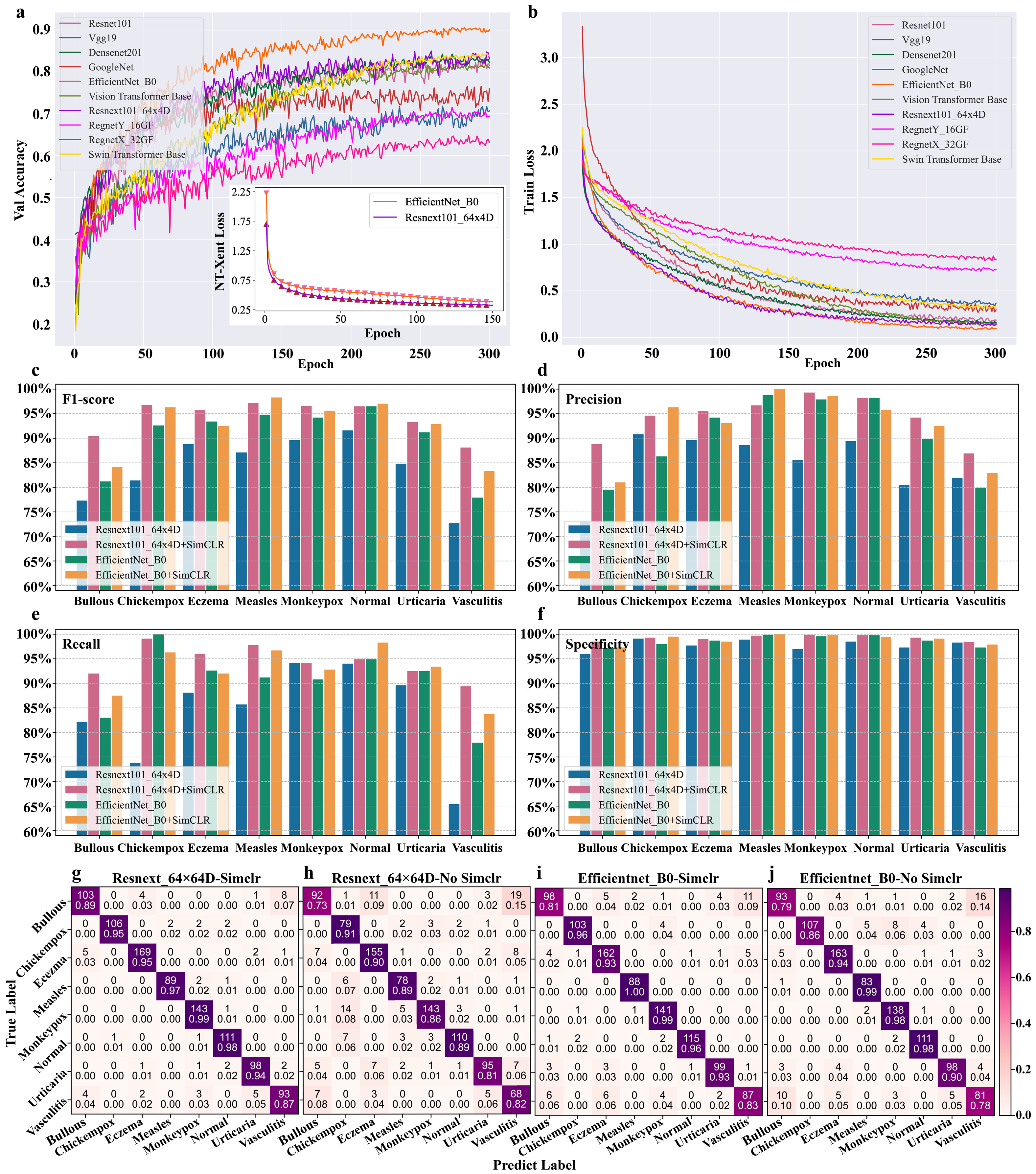}
    \caption{a,b: Test Accuracy and Train loss trends for the ten models, respectively. c: F1-score for test set of Data\_C. d: Precision for test set of Data\_C. e: Recall for test set of Data\_C. f: Specificity for test set of Data\_C. g - j: Confusion matrix of Renext101\_64x4D with SimCLR, Renext101\_64x4D, Efficientnet\_B0 with SimCLR and Efficientnet\_B0 for test set of Data\_C.}
    \label{fig3}
\end{figure}
During the SSL, we adjusted the size of input images to 224 × 224 pixels and uniformly used stochastic gradient descent (SGD) to optimize model parameters. Data augmentation in SimCLR was set as random crop and resize + random color jitter. We recorded the NT-Xent Loss values after each epoch and plotted the trend of NT-Xent Loss (Fig. 3a). The experimental results indicate that the NT-Xent Loss of Resnext101\_64x4D is lower than that of EfficientNet\_B0 throughout the whole process, which signifies Resnext101\_64x4D benefits more from SSL when NT-Xent Loss is only compared. 

Data\_C was utilized to retrain and retest two models pre-trained by SimCLR. The experimental equipment and the hyperparameters of the training process were the same as before. Meanwhile, the models were comprehensively evaluated using five metrics, i.e., Accuracy, F1-score (Fi3. 4c), Precision (Fig. 3d), Recall (Fig. 3e), Specificity (Fig. 3f) and confusion matrix (Fig. 3g-j). Meanwhile, the above metrics were also used to evaluate the models without SimCLR to demonstrate the potential of SSL. In terms of Accuracy, Resnext101\_64x4D improved from 84·96\% previously to 94·51\% and EfficientNet\_B0 improved from 90·51\% previously to 92·5\%. In terms of F1-score (Fig. 3c), Resnext101\_64x4D pre-trained by SimCLR was the most advanced in six categories (i.e., Bullous, Chickenpox, Eczema, Mpox, Urticaria, and Vasculitis), reaching 90·4\%, 96·8\%, 95·7\%, 96·6\%, 93·3\%, and 88·1\% respectively. In terms of Precision (Fig. 3d), Resnext101\_64x4D pre-trained by SimCLR is the most advanced in the six categories (i.e., Bullous, Eczema, Mpox, Normal, Urticaria, and Vasculitis), reaching 88·8\%, 95·5\%, 99·3\%, 98·2\%, 94·2\%, and 86·9\% respectively. Regarding Recall (Fig. 3e), Resnext101\_64x4D pre-trained by SimCLR learning was the most advanced in five categories (i.e., Bullous, Eczema, Measles, Mpox, and Vasculitis), reaching 92·0\%, 96·0\%, 97·8\%, 94·1\%, and 89·4\% respectively. In terms of Specificity (Fig. 3f), Resnext101\_64x4D pre-trained by SimCLR was the most advanced in the six categories (i.e., Bullous, Eczema, Mpox, Normal, Urticaria, and Vasculitis), achieving 98·5\%, 99·0\%, 99·9\%, 99·8\%, 99·3\%, and 98·4\%, respectively. Consistent with our considerations, Resnext101\_64x4D benefited more from SSL and was superior on the confusion matrix (Fig. 3g-j). Based on the results of these metrics, Resnext101\_64x4D with SimCLR was eventually deployed in Mpox-AISM.

\subsection{Mpox-AISM Grading Assessment}
To enable Mpox-AISM to be conveniently deployed in various real-world settings, we, according to the characteristics of the dataset, the evolution trend of Mpox rash and clinical features of most cases, further classified images of Mpox rashes into four grades and evaluated the performance of the model in predicting these images (Fig. 4a-b): Grade-I (faces, necks and hands) is a body part that is not easily covered and has high incidence; Grade-II (arms and legs) is a body part that is easier to check; Grade-III (back and chest); These three grades photos were taken from a distance. Grade-IV, or Others, is taken at close range. After evaluating these images via Mpox-AISM, the Recall rates of Grade-I, Grade-II, Grade-III and Others are 98·59\%, 100·00\%, 94·59\% and 99·33\%, respectively.

Accurate diagnosis for earlier-stage mpox helps curb the epidemic's spread. However, the symptoms of Mpox rash are not severe in the earlier-stage and are easily confused with other skin diseases. So, we specially tested Mpox-AISM’s performance using rash images of earlier-stage Mpox patients (Fig. 4c). Experimental result shows that Mpox-AISM achieves 100\% Recall in testing the images of earlier-stage mpox patients and has excellent diagnosis performance in diagnosing the images of later-stage Mpox patients too. Most importantly, the Accuracy of the earlier-stage mpox means that even in the earlier-stage of Mpox rash, Mpox-AISM can also make a primary diagnosis of the suspected case.
\begin{figure}
    \centering
    \includegraphics[width=0.8\textwidth]{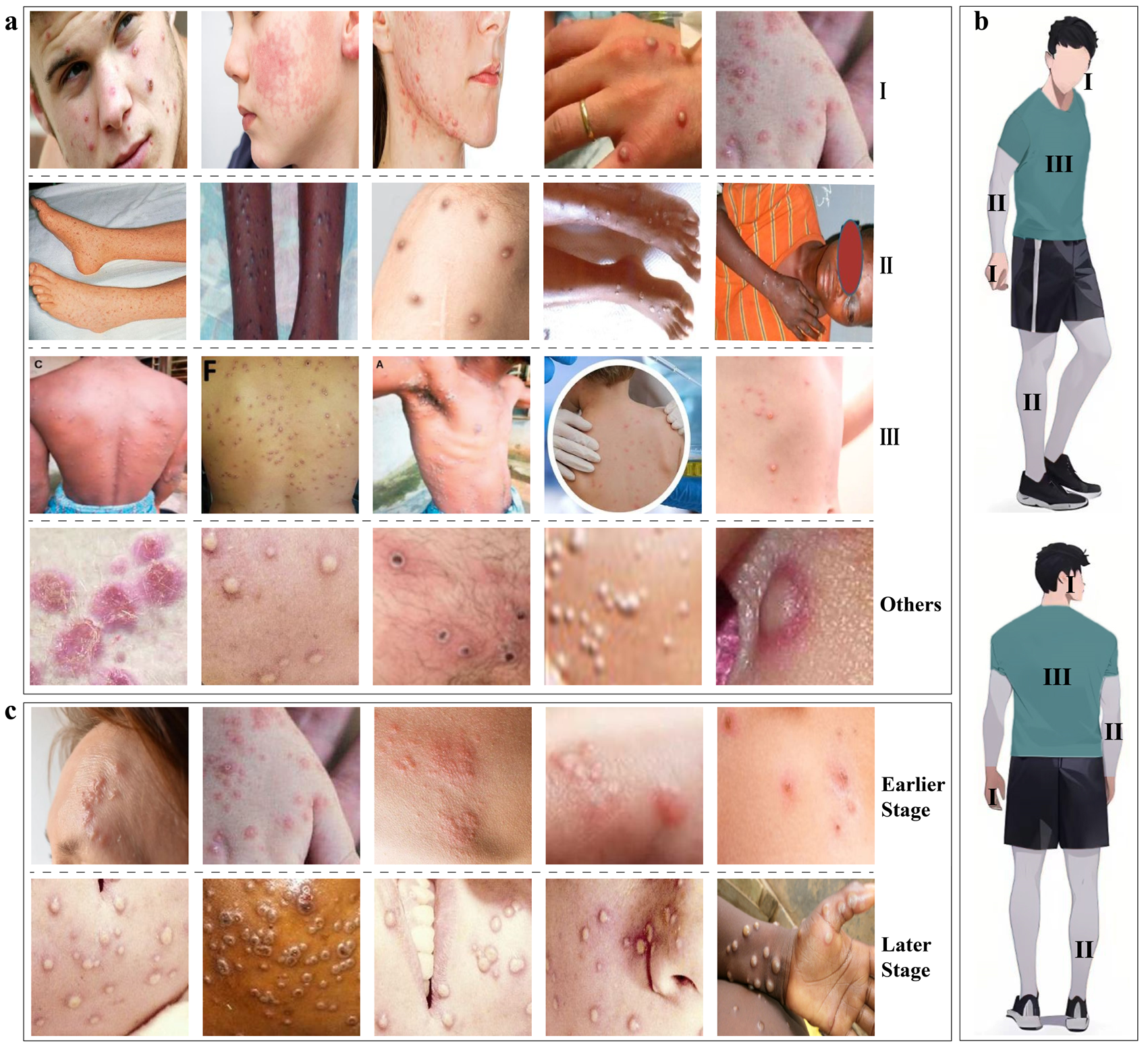}
    \caption{a: Diagrams of mpox rash (Grade-I, Grade-II, Grade-III and Others). b: Diagrams of human body parts. c: Diagrams of mpox rash at earlier-stage and later-stage.}
    \label{fig4}
\end{figure}

\subsection{Mpox-AISM Interpretability}
Deep learning models have exhibited superior performance in various tasks. However, due to their over-parameterized black-box nature, it is often difficult to understand the prediction results of deep models \cite{ref15}. The lack of interpretability raises a severe issue about the trust of deep models in high-stakes prediction applications, such as autonomous driving, healthcare, criminal justice, and financial services \cite{ref16}. Especially in the healthcare field, the results predicted by the models will affect the patient's subsequent treatment, so it is imperative to interpret these results. In addition, the WHO's Ethics and Governance of Artificial Intelligence for Health: WHO Guidance, published in 2021, specifies that AI should be intelligible or understandable to developers, users and regulators \cite{ref17}. Therefore, it is necessary to provide interpretable techniques in model prediction.
\begin{figure}
    \centering
    \includegraphics[width=0.8\textwidth]{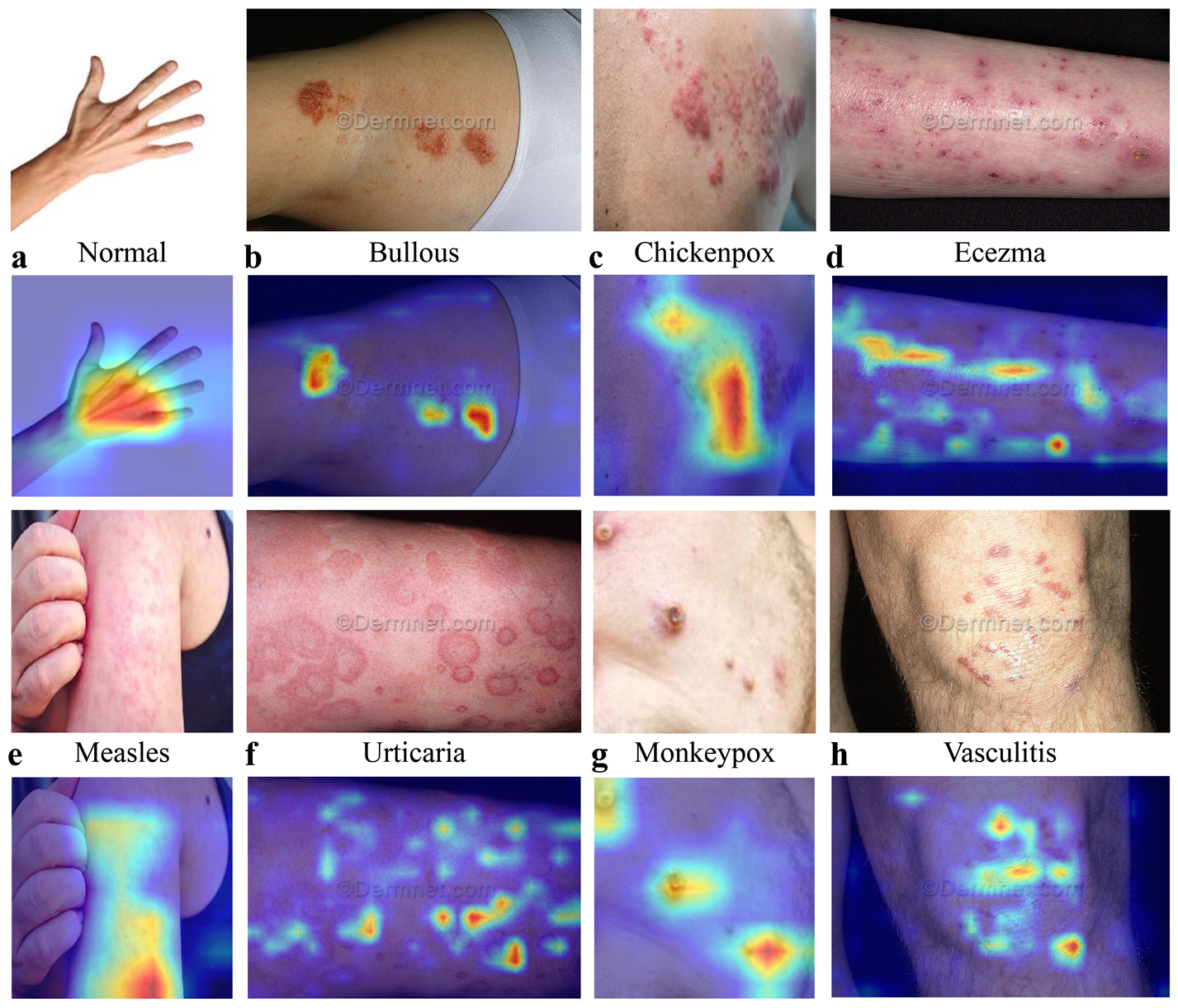}
    \caption{Heat maps for eight categories of skin diseases generated by the Grad-CAM method.}
    \label{fig5}
\end{figure}
In our work, we employed Gradient-weighted Class Activation Mapping (Grad-CAM) to explain the Mpox-AISM's decision-making process. Grad-CAM is a result visualization and interpretation technique that makes prediction results made by deep learning models more transparent. Grad-CAM helps in visualizing the regions of the image that are important for a particular classification. The gradient information is used to calculate the activation map of CNN for the input image, and the magnitude of the activation map can indicate the degree of influence of the image classification result on each part of the original image. Fig. 5 shows the heat maps of the diagnosis results generated by the Grad-CAM method where red stands for high relevance, yellow stands for medium relevance, and blue stands for low relevance. From Fig. 5, it can be seen that our model focuses well on the lesion region.
\subsection{Mpox-AISM Application}
We designed PC (Fig. 6b) and mobile (Fig. 6a) application pages corresponding to Mpox-AISM for ease of use. The PC terminal combines the terminal camera to capture the target image for diagnosis, which can be applied to entry-exit inspection in airports and customs (Fig. 1b). Mobile terminals, such as mobile phones, users only need to upload skin images from mobile phone lens or album by clicking the button located at the center of the screen, and then the categories of skin lesion area can be predicted, which provide a primary diagnosis to the user. The mobile terminal can be applied to family doctors, rural areas in underdeveloped regions, the wild and other settings (Fig. 1c-e). The Mpox-AISM corresponding terminal in this study is highly convenient and imposes no strict restrictions on the operator's photographing angle or distance thanks to multiple data augmentation strategies. 

To further improve the reliability of the application system, we carried out prediction probability distribution statistics on the test set (Illustration in Fig. 6c). It was found that in the test set, the proportion of samples with prediction probability $\geq$ 0·6 is 94\%, and the Accuracy of Mpox-AISM for these samples is almost 95·9\%. For Mpox images in these samples, Precision, Recall, Specificity and F1-score achieve respectively 99·3\%, 95·9\%, 99·9\% and 97·6\%. Besides, the proportion of samples with a prediction probability $>$ 0·5 is 5·4\%, and the error is $<$ 98\%. So, we set the prediction threshold at the application terminal to 0·6. If the application displays a case with a prediction probability value less than 0·6, the application will give a prompt that manual intervention is required. In addition, we tested the model using ten images of our normal skin and images (23 bullous images, 13 eczema images, two chickenpox images, two measles images, eight vasculitis images and 12 urticaria images) of 60 cases in the dermatology atlas. The results show that the prediction results of the model are almost consistent with the ground truth. That is, the model accuracy achieved 93\%.

Notably, the application also provided confidence in results and typical pathological images of various parts to improve the interpretability of results and the vigilance of patients. Our application can help suspected patients and doctors preliminarily screen and diagnose lesion areas anytime and anywhere without cost. In particular, during the outbreak of Mpox, such applications can provide specific technical support for limiting the spread of the epidemic.
\begin{figure}
    \centering
    \includegraphics[width=0.8\textwidth]{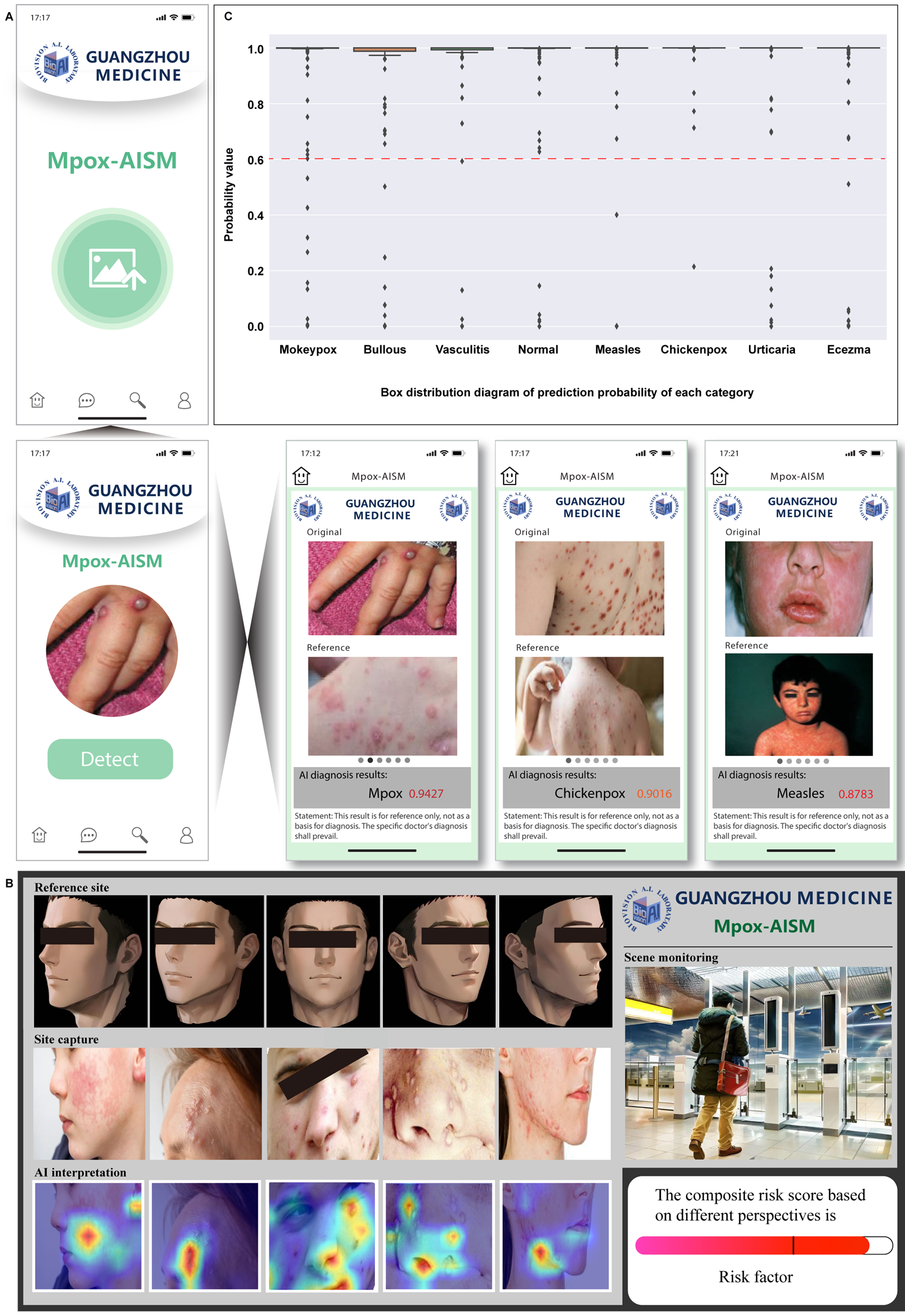}
    \caption{a: Mobile application page. b: PC application page. c: Box distribution diagram of prediction probability of each category.}
    \label{fig6}
\end{figure}
\section{Discussion}

In this study, we aimed to develop an efficient and real-time mpox diagnosis technology to help control the spread of mpox. In view of the unique appearance characteristics of mpox, we proposed Mpox-AISM, an innovative diagnosis strategy integrating AI and Internet technology. With the help of the Internet, Mpox-AISM can swiftly realize the visual diagnosis of mpox through the captured a skin rash image and distinguish it from six common like-mpox skin diseases and normal skin. The widely recognized public dataset was utilized to verify Mpox-AISM. The experimental results showed that the accuracy, precision, recall, specificity, and F1-score of Mpox-AISM reached 94.56\%, 99.3\%, 94.1\%, 99.9\%, and 96.6\%, respectively. Especially in the diagnosis of mpox, the precision, recall, specificity, and F1-score of Mpox-AISM reached 99.3\%, 94.1\%, 99.9\%, and 96.6\%, respectively, showing its excellent diagnosis performance.

The rapidly developing AI has made significant contributions to simplifying clinical processes and decision-making in health care \cite{ref11}. However, the performance of an AI model largely depends on the quality of the data set \cite{ref18}. For healthcare professionals, collecting a large number of medical images with accurate labels is facing great challenges, while obtaining unlabeled medical images is relatively easy \cite{ref19}. Especially for diseases such as mpox, it is particularly difficult to establish a high-quality dataset, because it requires dermatologists to spend much time on image acquisition and the accurate diagnosis of skin lesions in numerous patients. Facing this challenge, we employed data augmentation technology to simulate the image changes that the model may encounter in real-world settings. It is worth noting that we have introduced self-supervised learning, an innovative unsupervised learning paradigm aimed at enhancing the performance of the deep learning model in diagnostic tasks. Researchers generally believe that the use of SSL in medical image processing is of great significance because it can build proxy tasks to represent and learn large-scale unlabeled data, thereby effectively improving the performance of downstream tasks (such as image classification) \cite{ref20}. Our experimental results show that the performance of the deep learning model using SSL in diagnosing various types of images has been significantly improved, which is consistent with the previous research results \cite{ref21,ref22,ref23}. Due to the implementation of SSL, we can also use a large number of captured unlabeled data to quickly iterate and optimize the deep learning model when Mpox-AISM is used to diagnose patients with skin rashes in real-world settings \cite{ref24}. To our knowledge, we adopted the SSL strategy in mpox diagnosis research for the first time and verified its effectiveness.

The outbreak of infectious diseases may spread rapidly around the world and bring huge risks to global public health \cite{ref25}. Consequently, enhancing the management of infectious diseases is crucial for preventing infection and mitigating associated risks. Research shows that mHealth technology can assist people in better detecting, monitoring, and managing infectious diseases, thus facilitating the rapid identification of potential epidemics \cite{ref26}. The World Health Organization defines mHealth as the medical and public health practice supported by mobile devices. Statistical data reveal that approximately 2.5 billion individuals globally possess smartphones, and about 4.9 billion individuals have access to the Internet \cite{ref27,ref28}. Simultaneously, advancements in optical technology, materials, and software engineering are making smartphones increasingly compact and powerful. Smartphone-based mHealth applications hold significant potential in facilitating unparalleled professional clinical diagnoses and treatments \cite{ref27}. Particularly, AI enables smart tools (such as smartphones) to assist primary healthcare providers in helping with a rough screening at the doorstep and in peripheral areas having poor doctor-patient ratios [11]. In light of these considerations, we have developed an Internet-enabled smartphone application (Mpox-AISM App) based on the proposed Mpox-AISM strategy. Given the widespread adoption of smartphones and the Internet, the Mpox-AISM App could be pivotal in maintaining public health safety and controlling the mpox outbreak. Despite the application's functions (reading images, sending images to the cloud server, and displaying diagnostic results to users) being straightforward and fundamental, it serves as an invaluable resource for both the public and primary healthcare providers in high-risk areas with limited medical resources. The Mpox-AISM app not only accelerates the preliminary screening of suspected cases and reduces reliance on professional medical facilities but also offers a dependable and accessible tool for healthcare providers and the public to manage the mpox outbreak more effectively. In the future, the Mpox-AISM app can aid individuals in high-risk areas and primary healthcare providers lacking specialized knowledge (such as community doctors, general practitioners, rural doctors, and family doctors) in swiftly conducting preliminary screenings for skin rashes, thereby urging at-risk individuals to seek professional medical care promptly and reducing the risk of transmitting the mpox virus. The Mpox-AISM app can significantly enhance the speed and efficiency of public health responses, particularly in communities requiring rapid diagnosis and response, thereby combating the spread of mpox, safeguarding public health, and curbing the outbreak.

The early diagnosis of mpox patients holds critical importance, as it not only markedly enhances the effectiveness of treatment and mitigates the long-term impact on the patient's health but also significantly curtails the speed of disease transmission \cite{ref29}. Building upon this foundation, we specifically evaluated the potential of Mpox-AISM in diagnosing earlier-stage mpox rash. This endeavor aims to enhance the confidence of users affected by earlier-stage mpox in Mpox-AISM. Clinically, mpox rashes, with significant morphological differences, can manifest across various body parts. Additionally, when deploying systems based on Mpox-AISM in public spaces, these systems are typically restricted to capturing individuals' facial and hand areas. Therefore, a grading assessment strategy has been implemented. Our dataset is divided into mutually exclusive subsets based on body parts, and the model subsequently predicts each subset. Using this evaluation strategy affords users comprehensive insight into the model's performance in diagnosing rashes across various body parts. The significance of interpretable AI in medical image processing has grown, as it not only bolsters the trust and comprehension of medical professionals in the AI decision-making process but also enhances the model's transparency and reliability. Here, we employed Grad-CAM (Gradient-weighted Class Activation Mapping) to highlight the areas in rash images that contribute most significantly to the model's predictions, thus offering an intuitive visual explanation for the model's decision-making process. The reason for choosing grad cam is that compared with other interpretable AI technologies, grad cam can be applied without modifying the model architecture and can be applied to various deep learning models \cite{ref30}.

At the outset of this work, numerous studies have successfully employed AI and rash images to diagnose mpox swiftly. We conducted a comprehensive review of these studies, detailing their findings and the advantages of our research in Table 1. Our analysis revealed that despite progress in previous research, numerous limitations persist. Specifically, previous studies predominantly focus on differentiating mpox from non-mpox cases, as well as diagnosing mpox, chickenpox, measles, and normal skin conditions. Indeed, an efficient model ought not only to diagnose mpox accurately but also to identify various non-mpox rashes accurately, thereby offering detailed preliminary screening results for the public in areas with limited medical resources and alleviating the workload of clinicians. Furthermore, from a clinical standpoint, the model's efficacy in the early diagnosis of mpox is crucial for controlling the outbreak. However, previous studies have overlooked evaluating the model's diagnostic capabilities for earlier-stage mpox. Additionally, most previous studies only developed and validated deep learning models, neglecting the development of corresponding applications. The most crucial thing is that all previous work only used data augmentation and supervised learning strategies to address the challenge of insufficient medical images, which may lead to the following drawbacks: 1) model performance depends on the number of annotated images used; 2) The model is susceptible to the impact of imbalanced datasets \cite{ref31}; 3) Overuse of data augmentation may lead to overfitting of the model. In contrast, our work effectively addresses the aforementioned shortcomings \cite{ref32}. Firstly, our model can not only diagnose mpox, chickenpox, measles, and normal skin but also diagnose four common skin diseases in daily life: eczema, urticaria, bullae, and vasculitis. Therefore, the Mpox-AISM App can serve as a skin disease screening tool for most patients with rashes and primary healthcare providers. Secondly, we specifically evaluated the performance of the model in diagnosis earlier-stage mpox and developed corresponding networked smartphone applications. Most importantly, for the first time, we adopted a joint strategy based on self-supervised learning and supervised learning to develop the model. We not only demonstrated for the first time that self-supervised learning can effectively improve the performance of mpox diagnostic models but also trained high-performance diagnostic models using a small number of labeled images.

\section{Limitations of study}
We acknowledge that this study has several limitations. Firstly, although data augmentation and self-supervised learning strategies have alleviated the model performance degradation caused by insufficient images, the diversity of our dataset still needs to be strengthened. Therefore, future research will seek to collaborate with professional institutions to obtain more clinical images to improve the diagnostic performance and robustness of Mpox-AISM. Secondly, the model used by Mpox-AISM has many parameters, which puts significant pressure on the computing resources of cloud servers. Therefore, future research will also focus on developing lightweight models to improve response speed and reduce computational costs. Thirdly, although Mpox AISM can recognize seven types of skin diseases and normal skin, it is still necessary to expand more disease categories in the dataset to improve its practicality. Finally, conducting a thorough clinical evaluation of AI algorithms before adopting them in practice is crucial \cite{ref43}. Therefore, further prospective clinical testing is needed to ensure its reliability and safety before the widespread use of the Mpox-AISM app. Given that mpox is an infectious disease, it is recommended that primary healthcare providers combine epidemiological survey results when using the Mpox-AISM app to make more accurate diagnoses and guide patients to refer them to professional institutions \cite{ref44}. In addition, considering that models based on multimodal inputs have better learning ability compared to single modal input models \cite{ref45}, we are planning to develop a multimodal model that combines rash images and epidemiological survey results, aiming to improve the diagnostic accuracy and stability of Mpox-AISM.
\begin{table}[]
\caption{Comparison of our work with previous studies. ‘SSL’ and ‘SL’ mean self-supervised learning and supervised learning, respectively.}
\resizebox{1\textwidth}{!}{%
\begin{tabular}{@{}cccp{2cm}p{3cm}ccc@{}}
\toprule
\textbf{Method}                                                            & \textbf{Year} & \textbf{Strategy} & \textbf{Dataset Size}                                                                           & \textbf{Image Categories}                                                                                                    & \textbf{Earlier-stage Mpox} & \textbf{Accuracy} & \textbf{Application} \\ \midrule
AICOM-MP \cite{ref33}                                                                   & 2024          & SL                         & 6124 labeled images                                                                             & mpox, non-mpox                                                                                                               & ×                           & 96.3\%            & ×                    \\ \hline \\[5pt]
MobileNetV3-large \cite{ref34}                                                         & 2024          & SL                         & 400 labeled images                                                                              & mpox, chickenpox, acne, normal                                                                                               & ×                           & 88.2\%            & ×                    \\ \hline \\[5pt]
PoxNet22\cite{ref35}                                                                   & 2023          & SL                         & 3192 labeled images                                                                             & mpox, non-mpox                                                                                                               & ×                           & 100.0\%           & ×                    \\\hline \\[5pt]
DenseNet201\cite{ref36}                                                                & 2023          & SL                         & 1710 labeled images                                                                             & mpox, chickenpox, measles, normal                                                                                            & ×                           & 97.6\%            & ×                    \\\hline \\[5pt]
MobileNetV3-Small\cite{ref37}                                                          & 2023          & SL                         & 2056 labeled images                                                                             & mpox, non-mpox                                                                                                               & ×                           & 96.0\%            & ×                    \\\hline \\[5pt]
ResNet18  \cite{ref38}                                                                 & 2023          & SL                         & 3192 labeled images                                                                             & mpox, non-mpox                                                                                                               & ×                           & 99.5\%            & ×                    \\\hline \\[5pt]
Vision Transformer \cite{ref39}                                                        & 2023          & SL                         & 3192 labeled images                                                                             & mpox, non-mpox                                                                                                               & ×                           & 94.7\%            & ×                    \\\hline \\[5pt]
MonkeyNet      \cite{ref40}                                                            & 2023          & SL                         & 8689 labeled images                                                                             & mpox, chickenpox, measles,normal                                                                                             & ×                           & 93.2\%            & ×                    \\\hline \\[5pt]
Xception+Denseet169   \cite{ref41}                                                     & 2022          & SL                         & 1754 labeled images                                                                             & mpox, chickenpox, measles, normal                                                                                            & ×                           & 87.13\%           & ×                    \\\hline \\[5pt]
MobileNetV2     \cite{ref42}                                                           & 2022          & SL                         & 3192 labeled images                                                                             & mpox, non-mpox                                                                                                               & ×                           & 91.1\%            & \checkmark                    \\\hline \\[5pt]

\textbf{Mpox-AISM*}     & \textbf{2024}       & \textbf{SL+SSL}    & \textbf{25331 unlabeled images + 4831 labeled images}  & \textbf{mpox, chickenpox, measles, bullous, eczema, urticaria, vasculitis, normal} & \textbf{100.0\% (Recall)}   & \textbf{94.5\%}   & \textbf{\checkmark} \\ \bottomrule
\end{tabular}}
\end{table}
In conclusion, this work utilized AI technology and the Internet to develop an innovative monkeypox diagnosis strategy called Mpox-AISM successfully. By reading digital images captured by the lenses of networked devices, Mpox-AISM can not only accurately diagnose monkeypox, but also recognize common like-mpox skin diseases and normal skin conditions. It is worth emphasizing that Mpox-AISM has demonstrated excellent performance in diagnosing early monkeypox, providing strong technical support for controlling the monkeypox outbreak. The unique design of Mpox-AISM allows it to be widely deployed on everyday electronic devices, enabling rapid monkeypox screening in various real-world settings. In addition, the Mpox-AISM App, tailored for smartphones, has shown great potential for application, especially in environments with limited medical resources. In daily life, the Mpox-AISM App can serve as a skin disease management tool for the public and healthcare providers, thereby significantly reducing the risk of the monkeypox virus to public health safety.
\section{STAR METHODS}
\begin{table}[]
\centering
\caption{KEY RESOURCES TABLE}
\resizebox{1\textwidth}{!}{%
\begin{tabular}{@{}lll@{}}
\toprule
\textbf{REAGENT or RESOURCE} & \textbf{SOURCE} & \textbf{IDENTIFIER} \\ \midrule
\textbf{Deposited data}               &                 &                     \\ \midrule
MSID                             &    Bala et al. \cite{ref40}            &   \url{https://www.kaggle.com/datasets/dipuiucse/monkeypoxskinimagedataset}      \\
MSLD                             &   Ali et al. \cite{ref46,ref47}              &                    \url{https://www.kaggle.com/datasets/nafin59/monkeypox-skin-lesion-dataset} \\ 
Dermnet                             &    kaggle.com             &\url{https://www.kaggle.com/datasets/shubhamgoel27/dermnet}                   \\ 
ISIC 2019   &  Tschandl et al, Noel et al and Marc et al \cite{ref48,ref49,ref50}    
&\url{https://challenge.isic-archive.com/data/\#2019}              \\ 
Dermatology atlas                             &   DXY website              &    \url{https://www.dxy.cn/}                 \\ \midrule
 \textbf{Software and algorithms}                            &                 &                     \\ \midrule
 VGG                            &   Simonyan et al.\cite{ref51}              &     \url{https://github.com/pytorch/vision/blob/main/torchvision/models/vgg.py}                \\ 
GoogleNet                             &   Szegedy et al. \cite{ref52}             &  \url{https://github.com/pytorch/vision/blob/main/torchvision/models/googlenet.py}                   \\ 
ResNet                             &     He et al. \cite{ref53}

            &          \url{https://github.com/pytorch/vision/blob/main/torchvision/models/resnet.py}           \\ 
ResNeXt                             &    Xie et al.\cite{ref54}

             &        \url{https://github.com/pytorch/vision/blob/main/torchvision/models/resnet.py}             \\ 
DenseNet                             &     Huang et al.\cite{ref55}

            &      \url{https://github.com/pytorch/vision/blob/main/torchvision/models/densenet.py}               \\ 
EfficientNet                             &   Tan et al.\cite{ref56}

              &          \url{https://github.com/pytorch/vision/blob/main/torchvision/models/efficientnet.py}           \\ 
RegNet                             &        Radosavovic et al.\cite{ref57}

         &          \url{https://github.com/pytorch/vision/blob/main/torchvision/models/regnet.py}           \\ 
Vision Transformer                             &     Dosovitskiy et al.\cite{ref58}

            &         \url{https://github.com/pytorch/vision/blob/main/torchvision/models/vision\_transformer.py}            \\ 
Swin Transformer                             &   Liu et al.\cite{ref59}

              &          \url{https://github.com/pytorch/vision/blob/main/torchvision/models/swin\_transformer.py}           \\ 
SimCLR                             &        Chen et al. \cite{ref12}

         &           \url{https://docs.lightly.ai/self-supervised-learning/examples/simclr.html}          \\ 
Grad-CAM                            &        Selvaraju et al. \cite{ref60}

         &            \url{https://github.com/frgfm/torch-cam}         \\ 
Pytorch     &   Version 1.13.0     &   \url{https://pytorch.org/}  \\ 
Python                             &       Version 3.8.0          &    \url{https://www.python.org/}                 \\ \bottomrule
\end{tabular}}
\label{table2}
\end{table}
\subsection{Resource availability}
\subsubsection{Lead contact}
Further information and requests for resources should be directed to and will be fulfilled by the lead contact, Yubiao Yue (jiche2020@126.com).
\subsubsection{Materials availability}
This study did not generate new unique reagents ( \textbf{Table \ref{table2}}).
\subsubsection{Data and code availability}
The author Zhenzhang Li (zhenzhangli@gpnu.edu.cn) will provide data as reasonably requested. The code and demo video link are available at https://github.com/zhenzhang-li/Mpox-AISM.
\section{Method Details}
\subsection{Data usage}
In this study, we employed two datasets. One was named Data\_A (3100 images, eight categories). Data\_A includes Mpox (381 images), Measles (91 images), Chickenpox (107 images), Eczema (881 images), Urticaria (265 images), Bullous disease (561 images, Bullous for short), Vasculitis (521 images), and Normal (293 images). Mpox, Measles, Chickenpox and Normal skin images were obtained from the Monkeypox skin images dataset (MSID) \cite{ref40} and Monkeypox Skin Lesion Dataset (MSLD) \cite{ref46,ref47} . The remaining four categories, i.e., Eczema, Urticaria, Bullous, and Vasculitis, were obtained from the Dermnet. Fig. 7 shows the example for each of the eight disease categories. The other was named Data\_B (25331, eight categories). Data\_B includes Melanoma (4522), Melanocytic nevus (12875), Basal cell carcinoma (3323), Actinic keratosis (867), Benign keratosis (2624), Dermatofibroma (239), Vascular lesion (253), Squamous cell carcinoma (628). The Data\_B was from training data of ISIC 2019 and utilized in self-supervised learning. Generally, the dataset in self-supervised learning does not require labels. Consequently, we merged these eight categories of images and randomly shuffled them.

\begin{figure}
    \centering
    \includegraphics[width=0.8\textwidth]{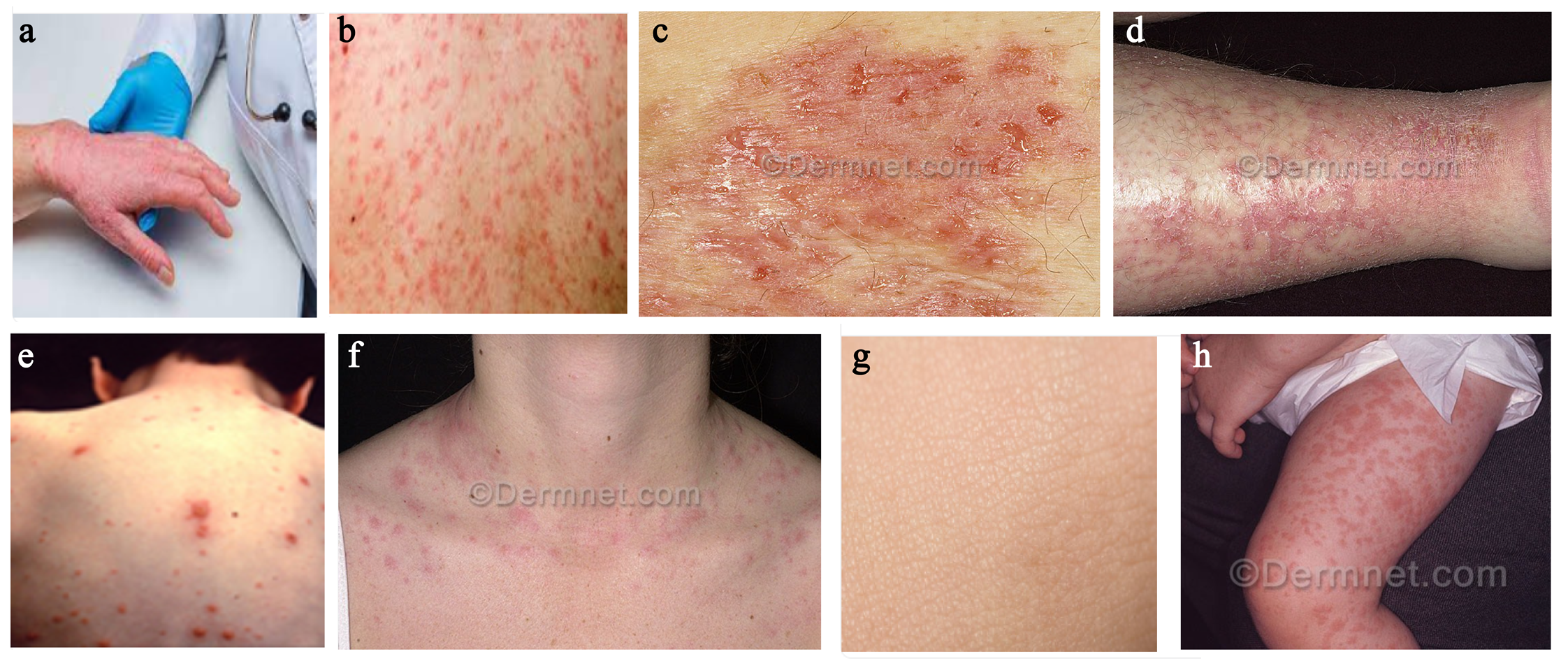}
    \caption{a: Mpox. b: Measles. c: Bullous. d: Eczema. e: Chickenpox. f: Urticaria. g: Normal. h: Vasculitis.}
    \label{fig:enter-label}
\end{figure}

\subsection{Data Augmentation, SimCLR and Evaluation Metrics}
Data is the driving force of deep learning, which determines the upper limit of models \cite{ref61}. Data augmentation alleviates the problem that insufficient samples hinder model performance by generating more data from limited data, enhancing the number and the diversity of samples, and improving model robustness\cite{ref62}. In the medical field, the dataset's insufficient sample size and category imbalance are especially prevailing\cite{ref63}. Therefore, it is necessary to perform an appropriate data augmentation strategy. In this study, we performed data augmentation for these five categories due to the scarcities of Mpox, Chickenpox, Measles, Normal and Urticaria. This work employed “Gaussian Noise + Crop and Resize + Affine + Cutout + Flip Horizontal + Flip Vertical + Gamma Contrast + Gaussian Blur (Random order and Random probability)”. Data\_A was expanded from the original 3100 to 4831 images by augmenting the above five categories. We labeled these 4831 images as Data\_C. Ultimately, the Data\_C was divided into a training set (3866) and a test set (965) in a ratio of 8:2. Here, the loss function of the models was uniformly set as Cross-Entropy-Loss, and the optimizer uniformly manipulated Stochastic Gradient Descent (SGD). This paper’s training and evaluation of models were performed in the Pytorch framework and an Ubuntu system with NVIDIA GeForce RTX 3090.
\begin{equation}
  {s_{i,j}} = \frac{{z_i^T{z_j}}}{{\tau \left\| {{z_i}} \right\|\left\| {{z_j}} \right\|}}  
\end{equation}

\begin{equation}
l(i,j) =  - \log \frac{{\exp ({s_{i,j}})}}{{\sum\limits_{k = 1}^{2N} {1[k! = i]\exp ({s_{i,k}})} }}
\end{equation}

\begin{equation}
 Loss = \frac{1}{{2N}}\sum\limits_{k = 1}^{2N} {\left[ {l(2k - 1,2k) + l(2k,2k - 1)} \right]}    
\end{equation}

When using SimCLR, the original input image is first randomly augmented twice, and then the new images generated are fed into the encoder simultaneously. Later, the encoder transforms the images into two vectors (hi, hj). Next, SimCLR employs a small neural network projection head to turn (hi, hj) into two new vectors (zi, zj). Finally, the NT-Xent-Loss between the two unknown vectors is calculated. Finally, the parameter information in the whole framework is updated with a configured optimizer according to the loss value. The NT-Xent loss was calculated as shown in Equation (1) to (3) and $\tau$ is a hyperparameter. 

\begin{equation}
Accuracy = \frac{{TP + TN}}{{TP + TN + FP + FN}}
\end{equation}

\begin{equation}
Precision = \frac{{TP}}{{TP + FP}}
\end{equation}

\begin{equation}
Recall = \frac{{TP}}{{TP + FN}}
\end{equation}

\begin{equation}
Specificity = \frac{{TN}}{{TN + FP}}
\end{equation}

\begin{equation}
F1\_score = \frac{{2Precision \times Recall}}{{Precision + Recall}}
\end{equation}
The objective evaluation for the model is essential. This article used five model evaluation metrics (Accuracy, Precision, Recall, F1-score, and Specificity), as shown in Equation (4) to (8). In these equations: TP means True Positive; TN means True Negative; FP means False Positive; FN means False Negative.

\section*{Acknowledgments}
This work was financially supported by the National Natural Science Foundation of China (Grant No\. 52172083), International Science \& Technology Cooperation Program of Guangdong (Grant No\. 2021A0505030078), and Guangzhou Key Research and Development Program (Grant No\. 2023B03J1239); Program for Innovative Research Team in University of Education System of Guangzhou (Grant No\. 202235404). 

\section*{Declaration of interests}
The authors declare no competing interests.
\bibliographystyle{unsrt}  
\bibliography{references}

\end{document}